\documentclass[superscriptaddress,secnumarabic,
amssymb,amsmath,nobibnotes,aps,prd,showkeys,showpacs,nofootinbib]{revtex4}%
\usepackage{graphicx}
\usepackage{epsf}
\usepackage{bm}
\usepackage{amsmath}
\usepackage{amsfonts}
\usepackage{amssymb}
\usepackage{epstopdf}
\usepackage{natbib}
\usepackage{color}
\providecommand{\U}[1]{\protect\rule{.1in}{.1in}}

\newcommand{\be}{\begin{equation}}
\newcommand{\ee}{\end{equation}}

\newcommand{\mincir}{\raise
-3.truept\hbox{\rlap{\hbox{$\sim$}}\raise4.truept\hbox{$<$}\ }}
\newcommand{\magcir}{\raise
-3.truept\hbox{\rlap{\hbox{$\sim$}}\raise4.truept\hbox{$>$}\ }}

\newtheorem{remark}{Remark}[section]

\begin{document}

\title{Different reheating mechanisms in quintessence inflation}

\author{Jaume  Haro\footnote{E-mail: jaime.haro@upc.edu}}
\affiliation{Departament de Matem\`atiques, Universitat Polit\`ecnica de Catalunya, Diagonal 647, 08028 Barcelona, Spain}

\thispagestyle{empty}

\begin{abstract}
Different  well-know ways to reheat the universe such as instant preheating,   the creation of particles nearly or conformally coupled with gravity,  or 
from the decay products of  a curvaton  field,
are revisited and discussed in detail in the framework of
  quintessence inflation, where the inflaton field at the end of inflation, instead to oscillate, rolls monotonically towards the infinite to drive the universe
to a kination regime.
For any kind of these preheating (particle creation) mechanisms, in order to calculate the reheating temperature, 
we point out the importance of the
Big Bang Nucleosynthesis bounds and 
the decay  process  of the massive fields involved in the theory, whose decay products
 form a relativistic plasma  whose energy density  eventually will dominate the one of the background after the phase transition.
\end{abstract}

\vspace{0.5cm}


\maketitle

\thispagestyle{empty}

\section{ Introduction}

A reheating  mechanism,  as pointed out by A. Guth in his seminal paper \cite{guth}, is an essential part of the inflationary paradigm in order match inflation with the hot Friedmann universe, because particles existing before the beginning of this period are completely diluted at the end of it. The most popular way to reheat the universe in standard inflation,
was via particle production  due to the oscillations of the inflaton field after the end of the inflationary period \cite{kls1}. Things changed after the discovery  of the current cosmic acceleration \cite{reiss, perlmutter} and the subsequent introduction of pioneering quintessence inflation models  to unify the early inflation period with the late time acceleration of the universe \cite{Spokoiny, pr, pv}, because the potential of these models 
 does not contain a deep well, and thus,  
the inflaton field rolls down without oscillations.  For these models, the various existing mechanisms of particle production  are completely different: 
The first one used in quintessence inflation was the gravitational particle production studied long time ago  in \cite{Parker,gm,glm,gmm} and at the end of nineties in \cite{Damour, Giovannini}, and applied to quintessence inflation in \cite{Spokoiny, pv, dimopoulos0, vardayan}. The second mechanism was the so-called {\it instant preheating} introduced in \cite{fkl0},  applied for the first time to quintessence inflation in \cite{fkl} {{} and recently in \cite{dimopoulos, vardayan} in the context of $\alpha$-attractors in supergravity.  The last one, that we study in this work,  is the {\it curvaton reheating} applied to brane-world inflation in \cite{LU} and to quintessence inflation in \cite{FL} and   \cite{ABM}. }
 
\

Dealing with the reheating temperature, the general convention is that it be
 the temperature of the universe at the beginning of the radiation-dominated era
as is mentioned in many papers (see for example the introduction of \cite{DLR, hannestad}). In standard inflation, i.e., when the potential contains a deep well, as we have already explained, the inflaton field begins  to oscillate creating a relativistic plasma of light particles, and assuming  as usual that they thermalize instantaneously, the reheating temperature coincides with the temperature of the universe when the inflaton field  decays in the minimum of the potential, what happens when the Hubble parameter is of the order of the decaying rate,
because all the energy density of the inflation vanishes when it reaches the minimum of the potential \cite{kls}. However, this does not happen
in quintessence inflation where the energy density of the inflaton field at the end of inflation does survive immediately after the phase transition,  which could lead to a misunderstanding in the definition of the reheating temperature.  For example,
 in curvaton reheating,   
the reheating temperature is sometimes calculated in the 
moment at which the curvaton field has completely decayed \cite{FL,ABM}) although the energy density of the curvaton field was subdominant at the end of its decay, or in instant preheating some authors identify
the reheating temperature as the temperature of produced particles when they are created, whose energy density is clearly subdominant at that moment \cite{sami, HMSS}.   Here we will use the usual convention, and we will denote by $T_R$ the temperature of the relativistic plasma in thermal equilibrium at the beginning of the radiation-domination era.

\

To obtain bounds for the reheating temperature, first of all one has to notice that the
radiation-domination era is previous to  the Big Bang  Nucleosynthesis (BBN) epoch which occurs in the  $1$ MeV  regime \cite{gkr}, meaning that the reheating temperature has to be greater
than $1$ MeV. 
On the other hand, 
 many supergravity and superstring theories contain particles, such as the gravitino or a modulus field, with only gravitational interactions, and consequently 
 the termal production of these relics and  its late time decay  may jeopardize the success of the standard BBN \cite{lindley}.
This problem can be solved by assuming sufficiently low reheating temperature, of the order of $10^9$ GeV \cite{eln}. Moreover,  in \cite{grt}  the authors 
realized,  using heuristic arguments, 
 that a non-thermal production of these gravitational relics 
 is possible during the inflationary phase, what  imposes upper bounds on the reheating temperature as low as $100$ GeV. In the present work, we will take a prudent viewpoint and we will restrict the reheating temperature to be 
 between $1$ MeV and $10^9$ GeV.

\

For some of the presented mechanisms, preheating
consists in the creation, after the phase transition to a kination regime,  of heavy massive particles that have to decay in lighter ones to form a relativistic plasma 
whose energy density, after  the plasma achieves the thermal equilibrium,  will eventually dominate in front of  the energy density of the background. This shows the importance of the parameters appearing in the  definition of the decaying rate, as the parameters involved in the particle production process such as the bare mass of the heavy particles, the coupling constant associated to the interaction between the inflaton and the quantum field,
the coupling constant associated to the interaction of the quantum field with gravity
 or the curvaton mass. All of them, must satisfy several algebraic constraints due to  some physical condition 
 such as,  the BBN success, the negligible effect  of the  vacuum polarization  during inflation or the fact that 
 the energy density of the massive produced particles was subdominant before its decay, imposed in order to obtain a viable reheating temperature.

\

The present work is structured as follows. In Section II we start discussing  the instant preheating formalism in non-oscillatory models showing that the particle production
is equivalent to  the Schwinger's effect. Once we have obtained the energy density of 
these massive particles, we study its decay in light products and find the constraints that must satisfy the parameters of the theory. For the viable values of these parameters we calculate the reheating temperature of the universe which is around a million of TeV. Section III is devoted  to the study of the    production of massless particles nearly conformally coupled with gravity, the
calculation of the vacuum modes is perturbative and the reheating temperature is in the TeV regime.
The production of heavy massive particles conformally coupled with gravity (particles with masses greater than the one of the inflaton) is considered in Section IV, where 
to calculate the vacuum modes one can use the WKB approximation. The reheating temperature is lower than in the previous cases, and depending on the values of the parameters involved in the decaying rate it could be in the GeV or TeV regime. Finally, in last Section we deal with the particle creation due to the decay of the curvaton field. We show that, in the case that  the curvaton decays when its energy density is subdominant,
in order to obtain temperatures compatibles with the BBN success the mass of the curvature field must be smaller than $10^{-7} M_{pl}$.


\

\

The units used throughout the paper are $\hbar=c=1$, and we denote  the Planck's and reduced Planck's mass respectively by $m_{pl}\equiv \frac{1}{\sqrt{ G}}\cong 1.2\times 10^{19}$ GeV and
$M_{pl}\equiv \frac{1}{\sqrt{8\pi G}}\cong 2.4\times 10^{18}$ GeV. We also denote, as we have already point out,  by $T_R$ the temperature of the universe at the beginning of the radiation epoch, which is the reheating temperature of the universe. Finally we will denote by $\Gamma$ the thermalization rate and by $\bar{\Gamma}$ the decay rate of a process.

\section{Instant Preheating in no-oscillatory models}
In this Section we review and discuss in detail the work of Felder,  Kofman and Linde \cite{fkl}  where
a mechanism called  {\it instant preheating}, which was   introduced in \cite{fkl0}  in the framework of standard inflation, was applied to the so-called   NO oscillatory models, i.e., to models where the inflaton field, instead to oscillate,  moves monotonically towards $\infty$.

\

Essentially,
instant preheating comes from  the interacting part of the Lagrangian density, namely $-\frac{1}{2}g^2\varphi^2\chi^2$, where 
g is a coupling constant,
$\varphi$ is the inflaton field and $\chi$ is a {{} a massive quantum field interacting with gravity} which,  due to a phase transition at the end of inflation, ceases to be in the vacuum state to produce  heavy massive particles. These created particles will
decay in light ones  forming  a relativistic plasma which will reach the thermal equilibrium and whose energy density eventually
dominate those of the background leading to a  radiation-dominated universe. {{} However, 
we want to stress that to obtain particle creation, instead of  considering an interaction between the inflation and que quantum field, one can deal with massive quantum field
with a
quartic self-interaction and coupled with gravity as has been recently done in \cite{tommi}. 
}

\

Once we have presented the idea behind   instant preheating, 
to understand the way that particles are produced it is useful to perform
 the change of variable $\bar{\chi}=a{\chi}$, to show that  the dynamical equation, in Fourier space, 
  for the $k$-mode of the quantum field $\chi$ becomes the equation of a time dependent harmonic oscillator
\begin{eqnarray}
\bar{\chi}_k''+\omega_k^2(\tau)\bar{\chi}_k=0,
\end{eqnarray}
where the derivative is with respect the conformal time $\tau$, and the square of the frequency is 
\begin{eqnarray}
\omega_k^2(\tau)=k^2+a^2(\tau)\left[m_{\chi}^2+g^2\varphi^2(\tau)+\left(\xi-\frac{1}{6}\right)R(\tau)\right],
\end{eqnarray}
being $m_{\chi}$  the bar mass of the quantum field,  $R(\tau)$  the scalar curvature and $\xi$  the coupling constant with gravity.

\

To clarify  and simplify ideas about preheating, we will  consider,  although the reasoning will serve for a general class of quintessence inflation models, the simple potential
\begin{eqnarray}\label{pot}
V(\varphi)=\left\{\begin{array}{ccc}
\frac{1}{2}m^2 \varphi^2 & \mbox{for}& \varphi<0\\
0 &\mbox{for}& \varphi\geq 0,
\end{array}\right.
\end{eqnarray}
which, at the end of inflation,  exhibits a phase transition to a kination \cite{Joyce} or deflationary \cite{Spokoiny} regime.

\

Note that, the inflationary part of the potential (\ref{pot}) is the well-known quadratic potential, which in the framework of  quintessence inflation where the number 
of e-folds is greater than in standard inflation due to the kination regime, provides a theoretical value of the spectral index, its running and 
a ratio of tensor to scalar perturbations matching with the recent observational data \cite{hap}. Moreover,  since
 after the phase transition the universe enters in a kination  regime, i.e.,
a phase that could be mimicked by and stiff fluid,  the energy density of the background will decay as $a^{-6}$, what allows that a relativistic plasma in thermal equilibrium, whose energy density evolves as $a^{-4}$, will eventually becomes dominant.

\

Choosing  $\tau=0$ as the phase transition time,  writing $\varphi(\tau)\cong \varphi'(0)\tau$ for values of $\tau$ close to zero, and considering a conformal coupling with gravity
$\xi=\frac{1}{6}$, 
we can approximate, near the phase transition,  the frequency $\omega_k(\tau)$ by
$\sqrt{k^2+a^2(0)(m_{\chi}^2+g^2(\varphi'(0))^2\tau^2)}$, where
approximating $a(\tau)$ by $a(0)$,  
we do not have taken into account the expansion of the
universe near the phase transition.

\

Taking in mind  all these approximations, what we have obtained is  the well-known over-barrier problem in scattering theory \cite{nikishov}, where the $\beta_k$-Bogoliubov coefficient is related with the reflection 
coefficient via the formula (see \cite{popov, marinov, haro} and \cite{meyer, fedoryuk} for a mathematical explanation)
\begin{eqnarray}
|\beta_k|^2=e^{-Im\int_{\gamma}\omega_{k}(\tau) d\tau},
\end{eqnarray}
being $\gamma$  a closed path in the complex plan containing the two turning points
$\tau_{\pm}=\pm i\frac{\sqrt{k^2+ a^2(0)m_{\chi}^2}}{a(0)g|\varphi'(0)|}.$

\

A simple calculation shows that the number of particles in the $k$-mode is given by \cite{haro}
\begin{eqnarray}\label{particle}
n_k\equiv |\beta_k|^2=e^{-\frac{\pi(k^2+a^2(0)m_{\chi}^2)}{a(0)g|\varphi'(0)|}}.
\end{eqnarray}

\

\begin{remark}
What is important in this approach is that, strictly speaking, $n_k$ will be the number of created particles at late times. However, as the authors argue  
in \cite{fkl}, 
high wave-length particles 
are produced when the effective mass of the field $\chi$ starts to change non-adiabatically, $g|\dot\varphi|\geq g^2\varphi^2$, what happens 
 at the phase transition. In fact, as one can see from formula (\ref{particle}), high frequency modes are exponentially suppressed.
\end{remark}

\

\begin{remark}
Essentially, instant preheating is the same as the Schwinger's effect \cite{schwinger}, i.e., the particle production in a constant electric field
\cite{nikishov1, bagrov, haro1}, because in both process the time dependent   
 frequency has the same particular form $\omega(t)=\sqrt{A^2+ B^2t^2}$, being $A$ and $B$  some constants.
\end{remark}

\

Then, the number density of produced particles is given, in terms of the cosmic time $t$, by  \cite{Birrell}
\begin{eqnarray}
n_{\chi}(t)\equiv \frac{1}{2\pi^2 a^3(t)}\int_0^{\infty} k^2|\beta_k|^2dk=
\frac{(g|\dot{\varphi}(0)|)^{\frac{3}{2}}}{8\pi^3}\left(\frac{a(0)}{a(t)}  \right)^3
e^{-\frac{\pi m_{\chi}^2}{g|\dot{\varphi}(0)|}},
\end{eqnarray}
where the dot means the derivative with respect the cosmic time, and  
 we have chosen $t=0$ as the phase transition
time. Analogously, the energy density of the produced particles is given by \cite{Birrell}
\begin{eqnarray}
\rho_{\chi}(t)\equiv \frac{1}{2\pi^2 a^4(t)}\int_0^{\infty}\omega_k(t) k^2|\beta_k|^2dk.
\end{eqnarray}

\

From these formulas, one can see that  at the phase transition 
\begin{eqnarray}
\rho_{\chi}(0)= \frac{g^2\dot{\varphi}^2(0)}{4\pi^4}
e^{-\frac{\pi m_{\chi}^2}{g|\dot{\varphi}(0)|}}                ,
\end{eqnarray}
and at late times  $\rho_{\chi}(t)\cong \sqrt{m_{\chi}^2+g^2\varphi^2(t)}n_{\chi}(t)$, meaning that  $\chi$-particles acquire a effective mass 
$m_{eff}\equiv \sqrt{m_{\chi}^2+ g^2\varphi^2(t)}$.

\

 Three  constraints must be imposed \cite{fkl}:
\begin{enumerate}\item
  To avoid  an exponential suppression of the energy density, the bare mass has to
 satisfy $m_{\chi}\leq \sqrt{g|\dot{\varphi}(0)|}$. For the sake of simplicity,  in this work we will take $m_{\chi}=0$.

 \item  Recalling that  the effective mass of the field $\chi$ is now  $g|\varphi|$ and taking into account that for masses  greater than the Hubble parameter the 
  vacuum polarization energy density 
  due to the field $\chi$, which 
  could be calculated using the WKB approximation, is of the order $\frac{H^6}{g^2\varphi^2} $  \cite{kaya}. 
Since  
  this
  quantity is   smaller than the energy density  of the background ($\sim H^2M_{pl}^2$), when there is a classical picture of the universe what happens for 
  $H\leq 10^{-2} M_{pl}$ (see Section IV), on can conclude that
  imposing  the condition $H\ll g|\varphi|$, 
  the polarization effects  will not affect the dynamics of the universe during inflation.  On the other hand, during the slow roll regime
$ H\cong \sqrt{\frac{V}{3M_{pl}^2}}$, thus for a quadratic potential the condition  $H\ll g|\varphi|$ is accomplished  imposing $g\gg \frac{m}{M_{pl}}$.

 \item The energy density of the produced $\chi$-particles cannot dominate before  decaying in light particles, which will form the   relativistic plasma,
 because if so,  the  force driving the inflation
back to $\varphi=0$ 
 will not disappear and the inflation field could not continue its movement forward up to $\infty$.
Effectively, the  interaction term $ \frac{1}{2}g^2\varphi^2\chi^2$ entails that  after the phase transition the inflaton field satisfies the equation
\begin{eqnarray}\label{inflaton}
\ddot{\varphi}+3H\dot{\varphi}=-g^2\chi^2\varphi.
\end{eqnarray}

Then, when the energy density of the $\chi$-particles is sub-dominant the right hand side of (\ref{inflaton}) is negligible and the field rolls towards $\infty$, but when 
it is dominant the right hand side of (\ref{inflaton}) ceases to be negligible, meaning that the inflaton field is under the action of  the quadratic potential $V(\varphi)=\frac{1}{2}g^2\varphi^2\chi^2$,
and thus,  it will roll down to zero, what may produce a new inflationary phase. 

\

To avoid this situation, first of all we have to calculate 
   the energy density of the background and the one of the field $\chi$ at the equilibrium, that  is,  when are of the same order 
\begin{eqnarray}
\rho(t_{eq})\sim \rho_{\chi}(t_{eq})\Longleftrightarrow 3H^2(t_{eq})M_{pl}^2\sim g\varphi(t)n_{\chi}(t_{eq}).
\end{eqnarray}

\

To obtain these quantities, first of all  we use that for  the model presented here,  whose potential is given by (\ref{pot}), after the phase transition the universe enters 
in a  kination regime and  the evolution of universe is given by
\begin{eqnarray}
\dot{H}=-3H^2 \Longleftrightarrow H(t)=\frac{H(0)}{3H(0)t+1} \Longleftrightarrow a(t)=a(0)(3H(0)t+1)^{\frac{1}{3}}.
\end{eqnarray}

Secondly,  using the Raychauduri equation $\dot{H}=-\frac{\dot{\varphi}^2}{2M_{pl}^2}$, one gets 
\begin{eqnarray}
\varphi=M_{pl}\int_0^t \sqrt{-2\dot{H}(s)}ds=M_{pl}\int_{H(t)}^{H(0)}\sqrt{\frac{-2}{\dot{H}(H)}}dH,
\end{eqnarray}
which for $t>0$, leads to
\begin{eqnarray}
\varphi(t)=\sqrt{\frac{2}{3}}M_{pl}\ln(3H(0)t+1).
\end{eqnarray}

\

Taking into account this results, and using that $\rho_{\chi}(t)\cong g\varphi (t) n_{\chi}(t)$, one  arrives at
\begin{eqnarray}
\rho_{\chi}(t)\cong \frac{6^{1/4}}{4\pi^3}g^{5/2}M_{pl}^{5/2}H^{3/2}(0)\ln(3H(0)t+1)
\left(\frac{a(0)}{a(t)} \right)^3,
\end{eqnarray}
and $\rho(t)=3H^2(0)M_{pl}^2\left(\frac{a(0)}{a(t)} \right)^6$. Then, both quantities are of the same order
when
\begin{eqnarray}
t\sim t_{eq}\equiv \frac{2\pi^3}{g^{5/2}\sqrt{H(0)M_{pl}}},
\end{eqnarray}
obtaining an important constraint for this theory: 
in order that  the back-reaction was subdominant and the inflaton field rolls monotonically towards  $\infty$,
the decaying time, i.e., when the particles has decayed in a relativistic plasma,  must be smaller than the equilibrium time $t_{eq}$.

\end{enumerate}

\

On the other hand,
assuming, as usual, that  there is no substantial drop of energy between the end of inflation and the phase transition time, 
and using that the value of the power spectrum of the curvature fluctuation in co-moving coordinates when the pivot scale leaves the Hubble radius is given by \cite{btw}
${\mathcal P}_{\zeta}\cong \frac{H_*^2}{8\pi^2 M_{pl}^2\epsilon_*}\sim 2\times 10^{-9}$ where $\epsilon=\frac{M_{pl}^2}{2}\left(\frac{V_{\varphi}}{V}\right)^2$ is the main slow roll parameter and the "star" means "when the pivot scale leaves the Hubble radius", one obtains 
\begin{eqnarray}
m^2\sim 3\times 10^{-9} \pi^2(1-n_s)^2   M_{pl}^2
\end{eqnarray}
 where we have used that for our model one has   $\epsilon_*=\frac{2M_{pl}^2}{\varphi^2_*}$ 
 and $\epsilon\cong \frac{1-n_s}{4}$, where $n_s$ denotes the spectral index. 
Then,  
  since  recent observations constraint the value of the spectral index to be $n_s=0.968\pm 0.006$ \cite{Planck}, taking its central value  one gets 
 $m\cong 5\times 10^{-6} M_{pl}$, and as a consequence, $H(0)\sim H_{end}\sim \frac{m\varphi_{end}}{\sqrt{6}M_{pl}}\cong 3\times10^{-6} M_{pl}$ (recall that inflation ends when 
  $\epsilon=1$). So, for values of $g \leq 10^{-2}$ one obtains $t_{eq}\geq 10^8 M_{pl}^{-1}$ and   $\varphi(t_{eq})\sim M_{pl}$. Thus,  for times  $t\in [10^{6} M_{pl}^{-1}, t_{eq}]$ the value of the inflation field remains close to $M_{pl}$, and the effective mass of the $\chi$-field will be 
   $gM_{pl}$.

\

 Now, 
 assuming that the $\chi$-field interacts with fermion particles, 
  the decaying rate will be
$\bar{\Gamma}=\frac{h^2 g \varphi(t)}{8\pi}$, where $h$ is a coupling constant (see \cite{lindebook} and references therein). As we have shown, the inflaton field spend most of time prior $t_{eq}$ at
$\varphi\sim M_{pl}$, then  one can safely take
$\bar{\Gamma}=\frac{h^2 g M_{pl}}{8\pi}$, and thus,
the condition $t_{dec}<t_{eq}$ , where $t_{dec}\cong \frac{8\pi}{3}\frac{1}{h^2g} M_{pl}^{-1}$ is the time when the field $\chi$ decayed, i.e.,  $H(t_{dec})\sim \bar{\Gamma}$, leads 
 to the relation 
$g< 3\times 10^{2}h^{4/3}$, which together with the condition $g\gg \frac{m}{M_{pl}}\cong 3\times 10^{-6}$, constraints the value of $h$ to fulfill
$h\gg  10^{-6}$.

\

Let's now calculate the  temperature at the equilibrium time assuming that previously the $\chi$-field has completely decayed, i.e., when $\rho(\bar{t}_{eq})\sim \rho_{\chi}(\bar{t}_{eq})$ (note that we  write $\bar{t}_{eq}$
instead of $t_{eq}$, because the equilibrium time is different depending on whether  the $\chi$-field decays or not).  In the case that $t_{dec}> 10^{6}M_{pl}^{-1}\Longrightarrow
h^2g<8\times 10^{-6}$, what means
$\varphi(t_{dec})\sim M_{pl}$,  since the corresponding energy densities evolve as
\begin{eqnarray}
\rho(\bar{t}_{eq})=\rho(t_{dec})\left( \frac{a(t_{dec}}{a(\bar{t}_{eq})}  \right)^6 \quad \mbox{ and } \quad 
\rho_{\chi}(\bar{t}_{eq})=\rho_{\chi}(t_{dec})\left( \frac{a(t_{dec}}{a(\bar{t}_{eq})}  \right)^4,
\end{eqnarray}
one has $\left( \frac{a(t_{dec})}{a(\bar{t}_{eq})}  \right)^2=\frac{\rho_{\chi}(t_{dec})}{\rho(t_{dec})}$, and thus, the temperature at the equilibrium is given by
\begin{eqnarray}
T_{eq}\sim \rho_{\chi}^{1/4}(\bar{t}_{eq})=\rho_{\chi}^{1/4}(t_{dec})\sqrt{ \frac{\rho_{\chi}(t_{dec})}{\rho(t_{dec})} }.
\end{eqnarray}

\

Taking into account that,
\begin{eqnarray}
\rho(t_{dec})= 3\bar{\Gamma}^2M_{pl}^2 \quad\mbox{and}\quad \rho_{\chi}(t_{dec})=gM_{pl}n_{\chi}(t_{dec})
\cong 10^{-2}g^{5/2}\sqrt{H(0)M_{pl}}M_{pl}^2\bar{\Gamma},
\end{eqnarray}
one gets
\begin{eqnarray}
T_{eq}\sim 2\times 10^{-2} g^{15/8} H^{3/8}(0) M_{pl}^{7/8}\bar{\Gamma}^{-1/4}
\cong 5\times 10^{14}\frac{g^{\frac{13}{8}}}{\sqrt{h}} \mbox{ GeV}.
\end{eqnarray}

\

In the case of a instantaneous thermalization, the reheating time $t_R$ coincides with $\bar{t}_{eq}$, and the reheating temperature will be 
$T_R=T_{eq}$
 what means that $ 10^{-11}h^{\frac{4}{13}}\leq g\leq  3\times 10^{-4} h^{\frac{4}{13}}$. Then, one can conclude that there is a narrow range of values of the parameters
 $h$ and $g$ accomplishing all the requirements:
 \begin{enumerate}
 \item $ 5\times 10^{-6}\ll g< 3\times 10^2 h^{4/3}$. (The back-reaction is not important during inflation and the $\chi$-field decays before equilibrium)
 \item $g< 8\times 10^{-6} h^{-2}$. (The decay ends when the value of the inflaton field is of the order of the reduced Planck mass).
 \item $ 10^{-11}h^{\frac{4}{13}}\leq g\leq  3\times 10^{-4} h^{\frac{4}{13}}$ (Reheating temperatures guaranteeing the BBN success).
 \end{enumerate}

Finally, the choice   ($h= 10^{-1}$,  $g= 10^{-4}$) or ($h=10^{-2}$, $g=5\times 10^{-5}$), which satisfy the above conditions, leads to 
  the same reheating temperature $T_R\sim 5\times 10^8$ GeV.

\

\section{Reheating via gravitational production of massless particles}
In this section,
we do not consider any interaction between the inflaton and the quantum field $\chi$, and we assume that the particles are nearly conformally coupled with gravity, i.e., 
$\xi\cong \frac{1}{6}$. Then,  the Klein-Gordon equation  for the quantum  $\chi$-field is given by
\begin{eqnarray}\label{a2}
\bar{\chi}''_{k}(\tau)+\left(k^2+
\left(\xi-\frac{1}{6}\right)a^2(\tau)R(\tau)\right)\bar{\chi}_{ k}(\tau)=0,
\end{eqnarray}
where once again $\bar{\chi}=a\chi$.

\

To define the vacuum modes before  and after the phase transition, we 
assume that
 at early and late time the term $a^2R$ will vanish, then the
 behavior at early and late times  is respectively
\begin{eqnarray}\label{vacuum}
 \bar\chi_{in, k}(\tau)\simeq \frac{e^{-ik\tau}}{\sqrt{2{k}}} (\mbox{ when }\tau\rightarrow -\infty), \quad \bar\chi_{out, k}(\tau)\simeq \frac{e^{-ik\tau}}{\sqrt{2{k}}}
 (\mbox{ when } \tau\rightarrow +\infty).
\end{eqnarray}

\

Therefore, 
 the vacuum modes at early ("in" modes) and late times
("out" modes) (exact solutions of (\ref{a2})) will be  given by \cite{ford}
\begin{eqnarray}\label{a36}
 \bar\chi_{in, k}(\tau)=\frac{e^{-ik\tau}}{\sqrt{2{k}}}-\frac{\xi-1/6}{{k}}
\int_{-\infty}^{\tau}a^2(\tau')R(\tau')\sin({ k}(\tau-\tau'))\bar\chi_{k}(\tau')d\tau', \nonumber\\
\bar\chi_{out, k}(\tau)=\frac{e^{-ik\tau}}{\sqrt{2{k}}}+\frac{\xi-1/6}{{k}}
\int_{\tau}^{\infty}a^2(\tau')R(\tau')\sin({ k}(\tau-\tau'))\bar\chi_{k}(\tau')d\tau'.
\end{eqnarray}

\

On the other hand,
since we are considering particles nearly conformally coupled {to gravity}, we can consider the term $(\xi-1/6)a^2(\tau)R(\tau)$ as a perturbation, and we can 
approximate the ``in''
and ``out'' modes
by the first order Picard's iteration, i.e., inserting (\ref{vacuum}) in the right hand side of (\ref{a36}),   {as}
\begin{eqnarray}\label{a37}
 \bar\chi_{in, k}(\tau)\cong \frac{e^{-ik\tau}}{\sqrt{2{k}}}-\frac{\xi-1/6}{{k}\sqrt{2{k}}}
\int_{-\infty}^{\tau}a^2(\tau')R(\tau')\sin({ k}(\tau-\tau')) e^{-ik\tau'} d\tau', \nonumber\\
\bar\chi_{out, k}(\tau)\cong \frac{e^{-ik\tau}}{\sqrt{2{k}}}+\frac{\xi-1/6}{{k}\sqrt{2{k}}}
\int_{\tau}^{\infty}a^2(\tau')R(\tau')\sin({ k}(\tau-\tau'))e^{-ik\tau'}d\tau',
\end{eqnarray}
which will represent, respectively, the vacuum before and after the phase transition.

\

After the phase transition, we could write the ``in'' mode as a linear combination of the ``out'' mode and its conjugate as follows
\begin{eqnarray}
 \bar\chi_{in,k}(\tau)=\alpha_k\bar\chi_{out,k}(\tau)+\beta_k\bar\chi^*_{out,k}(\tau),
\end{eqnarray}
and
imposing the continuity of $\bar\chi$ and its first derivative at the transition time we obtain,
up to order $\left(\xi-1/6  \right)^2 $,
the value of these
coefficients \cite{Birrell1, Zeldovich} will be
\begin{eqnarray}\label{bogoliubov}
  \alpha_k\cong 1-\frac{i({\xi}-\frac{1}{6})}{2k}\int_{-\infty}^{\infty}a^2(\tau)
 R(\tau) d\tau,\quad  \beta_k\cong \frac{i({\xi}-\frac{1}{6})}{2k}\int_{-\infty}^{\infty}e^{-2ik\tau}a^2(\tau)
 R(\tau) d\tau,
\end{eqnarray}
where
the integral of the $\beta$-Bogoliubov coefficient (\ref{bogoliubov}) is convergent because at early and late time, the term $a^2(\tau)R(\tau)$ converges fast enough 
to zero.

\

The energy density of the produced particles due to the phase transition is given by
\cite{Birrell}
\begin{eqnarray}\label{rho}
 \rho_{\chi}=\frac{1}{2\pi^2a^4}\int_0^{\infty}k^3|\beta_k|^2 dk,
\end{eqnarray}
where, if at the transition time, namely once again  $t=0$, the first derivative of the Hubble parameter is continuous one has
$\beta_k\sim {\mathcal O}(k^{-3})$, what means that
 this energy density is not ultra-violet divergent,  
and it approximately becomes \cite{ford}
\begin{eqnarray}\label{rho1}
 \rho_{\chi}(t) \cong \left({\xi}-\frac{1}{6}\right)^2{\mathcal N}H^{4}(0)\left(\frac{a(0)}{a(t)}\right)^4,
\end{eqnarray}
where ${\mathcal N}$ is a dimensionless numerical factor.

\

To understand this formula, we take $\tau=0$ as the value of the conformal time at the transition time, and we assume that at that time the second (or greater) derivative of the Hubble parameter is discontinuous. Then, since $\frac{d^n(a^2R)}{d\tau^n}(0)=C(n)a^{2+n}(0)H^{2+n}(0)$, where $C(n)$ is a dimensionless constant that only depends on $n$, integrating by parts one gets
\begin{eqnarray}
\beta_k=\left(\xi-\frac{1}{6}\right)\sum_{n=1}^{\infty}(-i)^nC(n)\frac{ a^{2+n}(0)H^{2+n}(0)}{(2k)^{2+n}}\equiv \left(\xi-\frac{1}{6}\right)f\left(\frac{k}{a(0)H(0)}\right),
\end{eqnarray}
where $f$ is some function. Thus, inserting this expression in (\ref{rho}) and performing the change of variable $s=\frac{k}{a(0)H(0)}$ on gets the expression 
(\ref{rho1}), with ${\mathcal N}=\frac{1}{2\pi^2}\int_0^{\infty}s^3|f(s)|^2ds$.

\begin{remark}
The number ${\mathcal N}$ is clearly model dependent. For the example proposed by Ford in \cite{ford} mimicking
a transition from de Sitter to a matter domination
modeled by $a^2(\tau)R(\tau)\equiv\frac{12}{\tau^2+\tau_0^2}$, $\mathcal  N$ could be calculated analytically giving as a result $\frac{9}{8}$.
However,  note that in this case reheating is impossible because the energy density of the produced particles decrease
faster that those of the background. We have calculated numerically this factor for some simple models that has a transition from a de Sitter regime to a deflationary one, and in all cases ${\mathcal N}$ is of the order 1 (see for instance \cite{ha}).
\end{remark}

\

{{}
Note also that, reheating via particle production of massless particles suffers the overproduction of gravitational waves \cite{pv} which could destabilize the  BBN process.  To avoid this challenge one has to impose that the {\it heating efficiency}, namely $\Theta$, and defined as the ratio of the energy density of the produced particles
 to the energy density of the background at the beginning of the kination epoch $\left(\Theta=\frac{\rho_{\chi}(0)}{\rho(0)}\right)$, satisfies the constraint \cite{rubio} (see also \cite{dimopoulos0})
\begin{eqnarray}
\Theta\geq 6\times 10^{-3}\left(\frac{H(0)}{M_{pl}} \right)^2\Longrightarrow \left(\xi-\frac{1}{6}\right)^2\geq 2\times 10^{-2}
\Longrightarrow \left(\xi-\frac{1}{6}\right)\sim 10^{-1}.
\end{eqnarray}

\

On the other hand, and
contrary to the previous Section, here}  we will  assume that thermal equilibrium of the produced particles is not an instantaneous process  \cite{allahverdi}. Instead of it, we 
will consider the following  thermalization rate   $\Gamma=n_{\chi}(0)\sigma_{2\rightarrow 3}$, where the most important process for the
chemical equilibrium are $2\rightarrow 3$ scatterings with gauge boson exchange  in the t-channel,  
whose typical energy energy is $E\sim H(0)\left(\frac{a(0)}{a(t)}\right)$ (see Section V of \cite{pv}), 
an the cross section is given by $\sigma_{2\rightarrow 3}=\frac{\alpha^3}{E^2}$ 
(see for instance the Section
IV of \cite{allahverdi}), where as usual, $\alpha^2\sim 10^{-3}$ \cite{Spokoiny}.

\

Since
\begin{eqnarray}
  n_{\chi}(t)=\frac{1}{2\pi^2 a^3(t)}\int_0^{\infty}k^2|\beta_k|^2 dk
  =\left({\xi}-\frac{1}{6}\right)^2{\mathcal M}H^{3}(0)\left(\frac{a(0)}{a(t)}\right)^3,
\end{eqnarray}
 where, for many models, one finds  \cite{hap1}{{}
\begin{eqnarray}{\mathcal M}\equiv \frac{1}{16\pi a^3(0)H^3(0)}\int_{-\infty}^{\infty} a^4(\tau)R^2(\tau) d\tau
\sim 1,
\end{eqnarray}
the thermalization  rate will acquire the form
\begin{eqnarray}
\Gamma=\alpha^3\left({\xi}-\frac{1}{6}\right)^2{\mathcal M}H(0)\left(\frac{a(0)}{a(t)}\right).
\end{eqnarray}

\

The relativistic fluid reach the thermal equilibrium  at $t=t_{th}$ when $H(t_{th})\sim \Gamma$ \cite{pv, Spokoiny}, i.e., for 
$\left(\frac{a(0)}{a(t_{th})}\right)^2\sim \alpha^3 \left({\xi}-\frac{1}{6}\right)^2{\mathcal M} $,
 meaning that the temperature at the thermalization time is of the order
\begin{eqnarray}
T_{th}\sim \rho_{\chi}^{1/4}(t_{th})\sim \alpha^{3/2}\left({\xi}-\frac{1}{6}\right)^{3/2}{\mathcal N}^{1/4}
{\mathcal M}^{1/2} H(0)\sim 
5\times 10^{-3}\left({\xi}-\frac{1}{6}\right)^{3/2}{\mathcal N}^{1/4}
{\mathcal M}^{1/2} H(0),
\end{eqnarray}
which for  typical values $H(0)\sim 10^{-6} M_{pl}$ and {{} ${\xi}-\frac{1}{6}\sim 10^{-1}$ leads to the temperature $T_{th} \sim 4\times 10^{8} \mbox{ GeV}$.}

\

 Finally, the equilibrium occurs  when the energy density of the background and the one of the created particles is of the same order ($\rho_{\chi}(t_{eq})\sim \rho(t_{eq})$),  what
 implies, for  $H(0)\sim  10^{-6} M_{pl}$, a temperature at the equilibrium of the order
 \begin{eqnarray}
 T_{eq}\sim \rho_{\chi}^{1/4}(t_{eq})=\rho_{\chi}^{1/4}(0)\sqrt{\frac{\rho_{\chi}(0)}{\rho(0)}}
 \sim \left({\xi}-\frac{1}{6}\right)^{3/2}{\mathcal N}^{3/4}
 \left(\frac{H(0)}{M_{pl}}\right)^2 M_{pl}
 \sim 2\times 10^{6}\left({\xi}-\frac{1}{6}\right)^{3/2}
 \mbox{GeV},
 \end{eqnarray}
which for {{} ${\xi}-\frac{1}{6}\sim 10^{-1}$ leads to the  temperature $T_{eq}\sim 6\times 10^{4}$ GeV.} Since $T_{eq}\leq T_{th}$,  one can conclude that
the thermalization occurs well before the equilibrium, and thus, the reheating temperature, i.e., the temperature of the universe when it is dominated by a 
relativistic plasma in thermal equilibrium,  is $T_R=T_{eq}$.

\

\section{Reheating via gravitational production of heavy massive particles}
In this section we will consider the creation of heavy massive particles conformally coupled with gravity, disregarding any interaction with the inflaton field, {{}
although it is also possible to deal with the production of heavy massive particles minimally coupled with gravity whose mass depends on the inflaton field \cite{rubio}}. In that situation, the frequency of the particles in the $k$-mode is $\omega_k(\tau)=\sqrt{k^2+m_{\chi}^2a^2(\tau)}$, and
during the adiabatic regimes,
 to calculate the $k$-vacuum mode, one can use the WKB approximation \cite{Haro}
\begin{eqnarray}
\bar{\chi}_{n,k}^{WKB}(\tau)\equiv
\sqrt{\frac{1}{2W_{n,k}(\tau)}}e^{-{i}\int^{\tau}W_{n,k}(\eta)d\eta},
\end{eqnarray}
where $n$ is the order of the approximation.

\

 When some high order derivatives
of the Hubble parameter are discontinuous, and thus the adiabatic regime breaks down, to obtain the evolution of the vacuum, one has to match the $k$-vacuum mode  
 (approximated by $\bar{\chi}_{n,k}^{WKB}$)  before this moment with a linear combination of positive and negative frequency modes 
 (approximated by a linear combination of $\bar{\chi}_{n,k}^{WKB}$ and its conjugate),
which is the manifestation  of the gravitational particle production. Basically, this  is  Parker's viewpoint of  particle creation in curved space-times  \cite{Parker}.

\

What is important  to keep in mind  is when 
it is possible to apply the WKB approximation. 
It is well-known that  at temperatures of the  order of the Planck's mass quantum effects become very important 
and it is impossible to have a classical picture of the universe. However, at temperatures below $m_{pl}=\frac{1}{\sqrt{G}}$, for example $T\sim 10^{-1}M_{pl}\sim 10^{17}$ GeV, 
as has been explained in the introduction of \cite{guth},  the beginning the  hot big bang scenario is possible.  Since, for the  flat FLRW universe 
$T\sim \rho^{1/4}\sim \sqrt{H M_{pl}}$ one 
can safely  deduce that 
a classical picture of the universe is possible at scales of the order  $H\sim 10^{-2}M_{pl}$. 
{{} Thus, at the beginning of inflation the Hubble parameter is practically constant and the universe is approximately in a de Sitter phase where, for a massive quantum field, the vacuum polarization, which was calculated in   \cite{bunch} showing that for masses less than the reduced Planck's one,    is subdominant with respect to the energy density of the background ($H^2M_{pl}^2$).
 If one wants that these polarization effects were also subdominant at the last stages of inflation, one could impose that  $m_{\chi}\gg H_*$, where
once again $H_*$ denotes the value of the Hubble parameter when the pivot scale leaves de Hubble radius, because one can use the WKB approximation to calculate the vacuum modes obtaining a vacuum energy density of the order $\frac{H^6}{m_{\chi}^2}$ \cite{kaya} which is also subdominant.
On the other hand, particles with mass greater than the Planck's one (in fact, for masses satisfying $m_{\chi}\geq \frac{m_{pl}}{\sqrt{2}}$) has a Compton wavelength smaller than its Schwarzschild radius 
$\frac{2 m_{\chi}}{m^{2}_{pl}}$. Then, these created particles becomes micro or Planck-size Black Holes, whose physics is unknown \cite{gt}
because the semiclassical thermodynamic description breaks down $-$ Hawking's formulas about evaporation are not applicable $-$ and it is not clear whether of how they radiate 
\cite{helfer}. For this reason, and since for a quadratic potential $H_*\cong 2\times 10^{-5} M_{pl}$, we have to consider massive quantum fields satisfying 
{}
\begin{eqnarray}
2\times 10^{-5}M_{pl}\ll  m_{\chi}\leq 2\sqrt{\pi} M_{pl}.
\end{eqnarray}
 }

\

In order to deal with an analytically solvable problem, i.e., having an analytic expression of the $\beta$-Bogoliubov coefficient, we consider a phase transition where the second derivative of the Hubble parameter is discontinuous, for example, the following model with an small cosmological
constant $\Lambda\sim 3H_0^2$, where $H_0$ is the current value of the Hubble parameter
\begin{eqnarray}\label{quadratic}
V(\varphi)=\left\{\begin{array}{ccc}
\frac{1}{2}m^2(\varphi^2-M_{pl}^2)+\Lambda M_{pl}^2 & \mbox{for} & \varphi\leq -M_{pl}\\
\Lambda M_{pl}^2 & \mbox{for} & \varphi\geq -M_{pl}.
\end{array}\right.
\end{eqnarray}

\

The effective Equation of State parameter is equal to $w_{eff}=-1-\frac{2\dot{H}}{3H^2}=-1+\frac{2}{3}\epsilon$, where 
$\epsilon=-\frac{\dot{H}}{H^2}\cong \frac{M_{pl}^2}{2}\left(\frac{V_{\varphi}}{V}  \right)^2$ is once again the mean slow-roll parameter.
For $\varphi\ll -M_{pl}$, one has $\epsilon\ll 1$ (slow-roll period) and then $w_{eff}\cong -1$. Immediately after the phase transition, since the cosmological constant has a very small value, all the energy is kinetic, and the universe enters in a kination \cite{Joyce} or deflationary \cite{Spokoiny} regime  with $w_{eff}\cong 1$. Finally, at the present time, due to the value of the cosmological constant, practically all the energy is potential, so $\dot{H}\cong 0$ and thus, $w_{eff}\cong -1$ showing the current cosmic acceleration.

\

Note that, the derivative of the potential is discontinuous at $\varphi=-M_{pl}$, what means, due to the conservation equation, that the second derivative of the 
inflaton field is discontinuous at the transition time, and consequently, form the Raychaudhury equation
$\dot{H}=-\frac{\dot{\varphi}^2}{2M_{pl}^2}$ one can deduce that the second derivative of the Hubble parameter is also discontinuous at this time.

\

In this case one only needs the  the first order WBK solution to approximate  the $k$-vacuum modes before and after the phase transition
\begin{eqnarray}
\bar{\chi}_{1,k}^{WKB}(\tau)\equiv
\sqrt{\frac{1}{2W_{1,k}(\tau)}}e^{-{i}\int^{\tau}W_{1,k}(\eta)d\eta},
\end{eqnarray}
where \cite{Winitzki}
\begin{eqnarray}
W_{1,k}=
\omega_k-\frac{1}{4}\frac{\omega''_{k}}{\omega^2_{k}}+\frac{3}{8}\frac{(\omega'_{k})^2}{\omega^3_{k}} ,
\end{eqnarray}
because $W_{1,k}$ contains the first derivative of the Hubble parameter, and since the matching involves the derivative of the mode, the $\beta$-Bogoliubov coefficient does not
vanish. 

\

Before the transition time, namely $\tau=0$, the  vacuum mode is depicted by $\chi_{1,k}^{WKB}(\tau)$, but after the phase transition this mode becomes a mix of positive and negative frequencies of the form
$\alpha_k \chi_{1,k}^{WKB}(\tau)+\beta_k (\chi_{1,k}^{WKB})^*(\tau)$.
The $\beta_k$-Bogoliubov coefficient is obtained matching both expressions at $\tau=0$, leading to
\begin{eqnarray}
\beta_k=\frac{{\mathcal W}[\chi_{1,k}^{WKB}(0^-),\chi_{1,k}^{WKB}(0^+)]}
{{\mathcal W}[(\chi_{1,k}^{WKB})^*(0^+),\chi_{1,k}^{WKB}(0^+)]},
\end{eqnarray}
where ${\mathcal W}[f(0^-),g(0^+)]=f(0^+)g'(0^-)-f'(0^+)g(0^-)$ is the Wronskian of the functions $f$ and $g$ at the transition time, and 
$F(0^{\pm})=\lim_{\tau\rightarrow 0}F(\pm|\tau|)$.

\

The square modulus of the $\beta$-Bogoliubov coefficient will be given approximately by   \cite{hap}
\begin{eqnarray}
 |\beta_k|^2\cong \frac{m^4_{\chi}a^{10}(0)\left(\ddot{H}(0^+)-\ddot{H}(0^-)\right)^2}{256(k^2+m_{\chi}^2a^2(0))^5},
\end{eqnarray}
with
\begin{eqnarray}
\ddot{H}(0^+)-\ddot{H}(0^-)=-\frac{\dot{\varphi}(0)}{M_{pl}^2}(\ddot{\varphi}(0^+)- \ddot{\varphi}(0^-))
=-\frac{\dot{\varphi}(0)}{M_{pl}^2}V_{\varphi}(-M_{pl}^-)= \frac{m^2\dot{\varphi}(0)}{M_{pl}}=
m^3,
 \end{eqnarray}
 where, assuming that  there is not subtantial drop of energy density between the end of inflation and the beginning of kination,  we have used that at the transition time, all the  energy at the end of inflation,  which is
  approximately $\frac{1}{2}m^2 M_{pl}^2$ because $\varphi_{end}=-\sqrt{2}M_{pl}$, was converted in kinetic.

\

Then, for our model, the number density of produced particles and its energy density
will be given by the following expressions
\begin{eqnarray}
 n_{\chi}(t)
 \sim 10^{-5}\left(\frac{m}{m_{\chi}}\right)^3 m^3\left(\frac{a(0)}{a(t)} \right)^3, \quad \rho_{\chi}(t)\sim m_{\chi}n_{\chi}(t).
\end{eqnarray}

{{}
\begin{remark}
Using the second order WKB approximation, the number density of produced particles is corrected by a term of the order $10^{-5}\left(\frac{m}{m_{\chi}}\right)^5 m^3$ whose contribution is
negligible due to the fact that $m\ll m_{\chi}$.
\end{remark}

}

\begin{remark}
Contrary to the procedure used in \cite{hap, hap1}, we cannot take as a thermalization rate $\Gamma=n_{\chi}\sigma_{2\rightarrow 3}$, because the created particles are very massive
and this rate is only justified for light particles. Instead of it,  first of all these particles has to decay in lighter ones, which will interact by the  exchange of bosons to reach the thermal equilibrium. In this Section, to simplify the calculations we will consider an instantaneous thermalization and we only will take into account the decaying process.
\end{remark}

Considering the decay  of the  $\chi$-field in fermions ($\chi\rightarrow \psi\bar\psi$), the  rate will be 
$\bar{\Gamma}=\frac{h^2 m_{\chi}}{8\pi}$ \cite{lindebook}, and the energy density of the background and the one of the relativistic plasma, when the decay is finished, 
i.e.,
 when $\bar{\Gamma}\sim H(t_{dec})=H(0)\left(\frac{a(0)}{a(t_{dec})} \right)^3\cong
  \frac{m}{\sqrt{6}}\left(\frac{a(0)}{a(t_{dec})}\right)^3 $, will be
\begin{eqnarray}
\rho(t_{dec})=3\bar{\Gamma}^2M_{pl}^2 \quad \mbox{and} \quad \rho_{\chi}(t_{dec})\sim 2\times 10^{-5}\left( \frac{m}{m_{\chi}} \right)^2 \frac{\bar{\Gamma}}{m}m^4.
\end{eqnarray}

\

{{}Imposing that the end of the decay is  before the domination of the relativistic plasma formed by the decay products $ \rho_{\chi}(t_{dec})\leq \rho(t_{dec})$}, one gets
\begin{eqnarray}\label{BOUND}
h^2\geq \frac{16\pi}{3}\times 10^{-5}\left( \frac{m}{m_{\chi}} \right)^3 \left(\frac{m}{M_{pl}}  \right)^2,
\end{eqnarray}
which for the value of the inflaton mass $m\sim 5\times 10^{-6}M_{pl}$ obtained in Section II, and taking the bare mass of the quantum field  {{}
one order greater than $H_*$, i.e., $m_{\chi}\cong 2\times 10^{-4}M_{pl}$, 
the
value of the coupling constant is constrained to satisfy $h\geq  2\times10^{-10}$.}

\

\begin{remark}
The end of the decay only happens after the domination of the relativistic plasma for abnormally small values of the parameter $h\leq 10^{-10}$. For this reason, here we will disregard  this situation.
\end{remark}

\

Then,
the reheating temperature, i.e., the temperature of the universe when the relativistic plasma in thermal equilibrium starts to dominate will be 
\begin{eqnarray}\label{reheating1}
 T_R\sim \rho_{\chi}^{\frac{1}{4}}(t_{dec})\sqrt{\frac{\rho_{\chi}(t_{dec})}{\rho(t_{dec})}} 
\sim 2\times 10^{-4}\left(\frac{m}{m_{\chi}}  \right)^{3/2}\left(\frac{m}{\bar\Gamma}  \right)^{1/4}\left(\frac{m}{M_{pl}}  \right)^2 M_{pl},
\end{eqnarray}
which for the above values of the inflaton mass and $m_{\chi}$, is of the order
{{}
\begin{eqnarray}
T_R\sim 7\times 10^{-19} h^{-1/2} M_{pl}\sim  h^{-1/2} \mbox{ GeV}.
\end{eqnarray}

 Taking  $h\sim 10^{-2}$, one gets reheating temperature  in the GeV regime, and  in the limit  case $h\sim 10^{-9}$, the reheating temperature would be 
 around $30$ TeV.
}

\

Now,  we calcule the reheating temperature for the following improved version of the well-known Peebles-Vilenkin potential \cite{pv}
\begin{eqnarray}\label{PV}
V(\varphi)=\left\{\begin{array}{ccc}
\frac{1}{2}m^2\left(\varphi^2-M_{pl}^2+M^2\right)& \mbox{for}& \varphi\leq -M_{pl}\\
\frac{1}{2}m^2\frac{M^6}{(\varphi+M_{pl})^4+M^4}& \mbox{for}& \varphi\geq -M_{pl},
\end{array} \right.
\end{eqnarray}
where $M$ is a parameter whose value could be calculated as follows:
 during the radiation and matter domination epoch the inflation field is all the time of
the order $M_{pl}$ (see \cite{pv} for a detailed discussion).
Then, in the model the field will dominate at late time when
\begin{eqnarray}\frac{m^2M^6}{M^4_{pl}+M^4}\sim \frac{m^2M^5}{M^4_{pl}}\sim H_0^2M_{pl}^2\Longrightarrow M\sim \left(\frac{H_0}{m} \right)^{\frac{1}{3}}M_{pl}\sim 
10^{-18} M_{pl}\sim 
1 \mbox{ GeV},
\end{eqnarray}
where we have used that the current value of the Hubble parameter is $H_0\sim 10^{-61} M_{pl}$.

\begin{remark}
The inflationary piece of  original Peebles-Vilenkin potential was quartic, and thus the theoretical values of  spectral index and the ratio of tensor to scalar perturbations do not enter in the marginalized  joint confidence contour in the plane  $(n_s,r)$ at $2\sigma$ C.L., without the presence of running \cite{hap}. For this reason, we have changed it by a quadratic potential, because the spectral values provided by a quadratic potential 
enter in this contour \cite{hap}.
\end{remark}

Since for this potential one also have  $\ddot{H}(0^+)-\ddot{H}(0^-)=m^3$, one can conclude that the reheating temperature will also be done by the formula (\ref{reheating1}).

\

To end the Section, and following the same spirit of the Peebles-Vilenkin model, we consider the following potential with a smoother phase transition than in the previous case
\begin{eqnarray}\label{PV1}
V(\varphi)=\left\{\begin{array}{ccc}
\frac{1}{2}m^2(\varphi^2+M^2)& \mbox{for}& \varphi\leq 0\\
\frac{1}{2}m^2\frac{M^6}{\varphi^4+M^4} &\mbox{for}& \varphi\geq 0.
\end{array}\right.
\end{eqnarray}

For this potential the discontinuity appears in the third derivative of the Hubble parameter, thus,  
 using the second order WKB approximation to obtain a non-vanishing $\beta$-Bogoliubov coefficient, one gets \cite{hap1}
\begin{eqnarray}
 |\beta_k|^2\cong \frac{m^4_{\chi}a^{12}(0)\left(\dddot{H}(0^+)-\dddot{H}(0^-)\right)^2}{1024(k^2+m_{\chi}^2a^2(0))^6},
\end{eqnarray}
where, taking into account that there is not a substantial drop of energy between the end of inflation and the phase transition to a kination regime, one has
\begin{eqnarray}
\dddot{H}(0^+)-\dddot{H}(0^-)=-\frac{\dot{\varphi}(0)}{M_{pl}^2}(\dddot{\varphi}(0^+)- \dddot{\varphi}(0^-))
=-\frac{\dot{\varphi}^2(0)}{M_{pl}^2}V_{\varphi\varphi}(0^-)= \frac{m^2\dot{\varphi}^2(0)}{M^2_{pl}}=2m^4.
\end{eqnarray}

\

The number density of massive produced particles and its energy density is given by 
\begin{eqnarray}
 n_{\chi}(t)
 \sim 8\times 10^{-6}\left(\frac{m}{m_{\chi}}\right)^5 m^3\left(\frac{a(0)}{a(t)} \right)^3, \quad \rho_{\chi}(t)\sim m_{\chi}n_{\chi}(t),
\end{eqnarray}
and for the same decaying rate as in the previous cases, 
the corresponding energy densities at the end of decay will be
\begin{eqnarray}
\rho(t_{dec})=3\bar{\Gamma}^2M_{pl}^2 \quad \mbox{ and } \quad \rho_{\chi}(t_{dec})\cong 10^{-5}\left(\frac{m}{m_{\chi}}  \right)^4\bar{\Gamma}m^3.
\end{eqnarray}

\

Assuming, once again,  that the end of the decay occurs  before the radiation-domination epoch ($ \rho_{\chi}(t_{dec})\leq \rho(t_{dec})$), one obtains the relation
\begin{eqnarray}\label{constraint0}
h^2\geq \frac{8\pi}{3}\times 10^{-5}\left(\frac{m}{m_{\chi}}  \right)^5\left(\frac{m}{M_{pl}}  \right)^2,
\end{eqnarray}
which for the values $m\cong 5\times 10^{-6} M_{pl}$ and {{} $m_{\chi}\cong 2\times 10^{-4} M_{pl}$ leads to the constraint $h\geq 3\times 10^{-12}$.}

\

Finally, if the thermalization of the relativistic plasma is instantaneous, the reheating temperature formula will be
\begin{eqnarray}
T_R\sim \rho_{\chi}^{1/4}(t_{dec})\sqrt{\frac{\rho_{\chi}(t_{dec})}{\rho(t_{dec})}}\sim 5\times 10^{-4} 
\left( \frac{m}{m_{\chi}} \right)^{\frac{13}{4}}\left( \frac{m}{M_{pl}} \right)^{2} h^{-1/2}M_{pl},
\end{eqnarray}
which for the masses as above, leads to the following  low reheating temperature
$T_R\sim 7\times10^{-20} h^{-1/2} M_{pl}\sim 10^{-1} h^{-1/2}$ GeV. From this result, one can conclude that 
for $h\sim 10^{-4}$ the reheating temperature is in the GeV regime, and to obtain a obtain a temperature in the TeV regime one needs $h\sim 10^{-8}$. The maximum
temperature is around $10$ TeV.


\

{{} An important final remark is in order: To avoid the problem of the overproduction of gravitational waves, following step by step the section V of \cite{rubio},  
in the case  of massive particle production whose energy density decreases  as $a^{-3}$ before decaying in a relativistic plasma (this process is previous to the equilibrium),  
the {\it heating efficiency} has to satisfy the constraint
\begin{eqnarray} \label{constraint}
\Theta \left(\frac{H(0)}{\bar{\Gamma}}  \right)^{1/3}\geq 6\times 10^{-3} \left(\frac{H(0)}{M_{pl}}  \right)^2.
\end{eqnarray}

When reheating is via instant preheating, there is no problem, because this constraint is satisfied for all the viable values of $h$ and $g$. Dealing with reheating via production of 
heavy massive particles,
 for the potentials (\ref{quadratic}) and (\ref{PV}), this constraint together with the bound (\ref{BOUND}) coming from the imposition that the decay was before 
the equilibrium, bounds the value of the mass of the quantum field to satisfy  $2\times 10^{-5} M_{pl}\ll m_{\chi}\leq 7\times 10^{-4} M_{pl}$,
what implies $m_{\chi}\sim 2\times 10^{-4} M_{pl}$
 and leads to a value of $h$ of the order 
$h\sim 10^{-9}$, obtaining a reheating temperature around $30$ TeV.
However, for the potential
(\ref{PV1}), the constraints (\ref{constraint}) and (\ref{constraint0}) leads to the bound $m_{\chi}\leq 10^{-5} M_{pl}$ which is incompatible with the bound
$m_{\chi}\gg 2\times 10^{-5} M_{pl}$,   
meaning that
in this case the amount  of created particles is not big enough to prevent that gravitational waves  influence the BBN process.}

}

\section{Curvature reheating in non-oscillatory models}
In this last Section we will review the so-called {\it curvature reheating} mechanism in quintessence inflation. To do that we will follow \cite{FL}, but taking into account that the authors of that paper, contrary to our convention, define the reheating temperature as the  temperature of the universe when the curvaton field has totally decayed in relativistic particles,  regardless of whether  its energy density  was dominant or not.

\

Assuming that the potencial of the curvaton field, namely $\sigma$,  is quadratic $V(\sigma)=\frac{1}{2}m_{\sigma}^2\sigma^2$, where
the mass of the curvaton is chosen 
to be  smaller  than the value of the Hubble parameter at the end of inflation $m_{\sigma}\ll H_{end}$. 
This choice ensures that  at the end of the inflation  the curvaton field was in a slow-roll regime,  because the 
condition $m_{\sigma}\ll H_{end}\Longleftrightarrow V_{\sigma\sigma}\ll H_{end}^2$  means that the curvaton potential is flat enough at that time \cite{LW}. Then, in order to avoid a second inflationary stage,  now driven by the curvaton, one has to impose that its energy density was subdominant when the curvaton starts to oscillate, what happens when $m_{\sigma}\cong H$  \cite{LW} (see also the section $5.5.1$ of \cite{mukhanovbook} for a detailed discussion of the quadratic potential).  Then,
\begin{eqnarray}
\rho_{\sigma}(t_{osc})<\rho(t_{osc})=3H^2(t_{osc})M_{pl}^2,
\end{eqnarray}
where $t_{osc}$ is the time when the curvaton starts to oscillate. 
Taking into account that $H(t_{osc})\cong m_{\sigma}$ and using the virial theorem, which for a quadratic potencial
states that the average over time of the kinetic and potential energy density coincide, 
we will make the approximation
$\rho_{\sigma}(t_{osc})\cong m_{\sigma}^2\sigma^2(t_{osc})$, obtaining the bound
$\sigma^2(t_{osc})<3 M_{pl}^2$.

\

Here,   the potential of the inflaton field, as the Peebles-Vilenkin one or the one used in Section II, is chosen under the requirement that after the end of inflation the universe enters in a kination regime. Which entails that, after the phase transition, the energy density of the background evolves as $a^{-6}$, and those of the curvaton as $a^{-3}$ because, during the oscillatory regime, the effective Equation of State parameter for a power law potential 
$V(\sigma)=V_0\left(\frac{\sigma}{M_{pl}}\right)^{2n}$ 
is given by $w_{eff}\cong \frac{n-1}{n+1}$ \cite{turner}.

\

Now, let $\bar{\Gamma}$ be the decay rate of the curvaton. There are two completely different situations: 
\begin{enumerate}
\item
The curvaton decays when it was subdominant.
\item 
 Te curvaton decays when the curvaton field dominated the universe.  
 \end{enumerate}
 
 In the first case, the curvaton decays in radiation (to simplify we assume the thermalization is instantaneous)  at a time ${t_{dec}}$ satisfying $H({t}_{dec})\cong \bar{\Gamma}$ (Note that the background  energy density is the one of the inflaton). At that time we will have
 \begin{eqnarray}
 \rho_{\sigma}({t}_{dec})<\rho({t}_{dec})\Longrightarrow \rho_{\sigma}(t_{osc})\frac{\bar{\Gamma}}{m_{\sigma}}<3\bar{\Gamma}^2M_{pl}^2,
\end{eqnarray}
where we have used that the energy density of the curvaton decays as $a^{-3}$, that the universe is in  a kination phase  (the Hubble parameter also decays as $a^{-3}$)  and $H(t_{osc})\cong m_{\sigma}$. Then, since 
$  \rho_{\sigma}(t_{osc})\cong  m_{\sigma}^2\sigma^2$ and $\bar{\Gamma}\cong H({t}_{dec})\leq H(t_{osc})\cong m_{\sigma}$ one gets the constraint
\begin{eqnarray}\label{bound}
\frac{\sigma^2(t_{osc})}{3M_{pl}^2}\leq \frac{\bar{\Gamma}}{m_{\sigma}}\leq 1.
\end{eqnarray}

\

To obtain the reheating temperature, which in this case coincides with the temperature at the  equilibrium time $t_{eq}$ ($\rho(t_{eq})\sim \rho_{\sigma}(t_{eq})$), one has 
to take into account that
\begin{eqnarray}
\rho(t_{eq})=\rho({t}_{dec})\left( \frac{a({t}_{dec})}{a(t_{eq})} \right)^6 \quad \mbox{and} \quad \rho_{\sigma}(t_{eq})=\rho_{\sigma}({t}_{dec})\left( \frac{a({t}_{dec})}{a(t_{eq})} \right)^4,
\end{eqnarray}
what leads to 
\begin{eqnarray}
T_R\sim \rho_{\sigma}^{1/4}(t_{eq})\sim \rho_{\sigma}^{1/4}({t}_{dec})\sqrt{\frac{\rho_{\sigma}({t}_{dec})}{\rho({t}_{dec})}}
\sim \frac{\rho_{\sigma}^{3/4}({t}_{dec})}{\sqrt{3}M_{pl}\bar{\Gamma}}
\sim \frac{m_{\sigma}^{3/4}|\sigma(t_{osc})|^{3/2}}{\sqrt{3}M_{pl}\bar{\Gamma}^{1/4}},
\end{eqnarray}
where we have used that $\rho_{\sigma}({t}_{dec})=\rho_{\sigma}(t_{osc})\left( \frac{a(t_{osc})}{a({t}_{dec})}\right)^3
=\rho_{\sigma}(t_{osc})\frac{H({t}_{dec})}{H(t_{osc})}\cong m_{\sigma} \sigma^2(t_{osc})\bar{\Gamma}$.

\

Then, using the bound (\ref{bound}) we can see that the reheating temperature is constrained to be in the range
\begin{eqnarray}\label{temperature}
\frac{m_{\sigma}^{1/2}|\sigma(t_{osc})|^{3/2}}{\sqrt{3}M_{pl}}\leq T_R\leq \frac{m_{\sigma}^{1/2}|\sigma(t_{osc})|}{{3}^{1/4}M^{1/2}_{pl}}.
\end{eqnarray}

\

On the other hand, when the decay of the curvaton occurs   when it is subdominant, the power spectrum of the curvature fluctuation in co-moving coordinates is given by \cite{LU, ABM}     
${\mathcal P}_{\zeta}= \frac{1}{1296\pi^2}\frac{m_{\sigma}^2}{\bar{\Gamma}^2} \frac{H_*^2\sigma_*^2}{M^4_{pl}}$, which from the bound (\ref{bound}), is constrained to satisfy
\begin{eqnarray}
\frac{1}{1296\pi^2} \frac{H_*^2\sigma_*^2}{M^4_{pl}}\leq {\mathcal P}_{\zeta} \leq \frac{1}{144\pi^2} \frac{H_*^2}{\sigma_*^2},
\end{eqnarray}
where we have used that before the oscillations the curvaton slowly rolls, and thus $\sigma(t_{osc})\sim \sigma_*$.
Now, taking into account the observational data
${\mathcal P}_{\zeta}\sim 2\times 10^{-9}$, one gets the bounds
\begin{eqnarray}
\frac{H_*}{|\sigma_*|}\geq  2 \times 10^{-3} \quad \mbox{ and } \quad {H_*}{|\sigma_*|}\leq  5 \times 10^{-3} M_{pl}^2. 
\end{eqnarray}

\

 Then, choosing $H_*\sim 2\times 10^{-5} M_{pl}$, which for 
 $n_s\cong 0.968$ and without the presence of the curvaton field,  is the value of the Hubble parameter when the pivot scale leaves the Hubble radius
 \cite{ha},  
one can safely {{} take $|\sigma_*|\sim 10^{-6} M_{pl}$, which agree with the bound $\sigma_* ^2\sim \sigma^2(t_{osc})<3M_{pl}^2$.}

\

For these values, the equation  (\ref{temperature}) becomes 
\begin{eqnarray}
5\times 10^{-9} \sqrt{m_{\sigma}M_{pl}} \leq T_R \leq 7\times 10^{-7} \sqrt{m_{\sigma}M_{pl}}, \end{eqnarray}
what means that, to get the bounds coming from BBN success,  one has to choose   curvaton masses satisfying
\begin{eqnarray}
  10^{-9} \mbox{ GeV } \cong 7\times 10^{-27} M_{pl} \leq m_{\sigma}\leq 3\times10^{-7} M_{pl}\cong 7\times 10^{11}  \mbox{ GeV},
\end{eqnarray}
and to avoid the effects of an overproduction of gravitational waves, taking into account that the energy density of the curvaton at the beginning of
the kination regime is approximately $m_{\sigma}^2 \sigma_*^2$,  the inequalities  (\ref{bound}) and (\ref{constraint}) leads to 
\begin{eqnarray}
\frac{m_{\sigma}\sigma_*^2}{3M_{pl}^2}\leq 2\times 10^5\frac{m_{\sigma}^6\sigma_*^6}{H^{11}(0)} \Longleftrightarrow 10^{-5}\leq 6 \frac{m_{\sigma}^5\sigma_*^4 
M_{pl}^2}{H^{11}(0)},
\end{eqnarray}
which is fulfilled for a wide range of viables values of the parameters. For example, choosing  $H(0)\sim 3\times 10^{-6} M_{pl}$, $\sigma_*\sim 10^{-6} M_{pl}$ and $m_{\sigma}\sim 10^{-7} M_{pl}$.

\

\

Finally,  assuming once again instantaneous thermalization, we consider the situation where the curvaton decays when dominates, that is, 
$\rho_{\varphi}({t}_{dec})\leq \rho({t}_{dec})$ (Now $\rho$ denotes the energy density of the curvaton), and the reheating will occur at $t_R\equiv\left(\frac{m}{m_{\chi}}\right)^3 m^3v {t}_{dec}$.
Let ${t}_{eq}$ be  once again the equilibrium time ($\rho_{\sigma}({t}_{eq})\sim \rho_{\varphi}({t}_{eq})$), which will satisfy $t_{osc}\leq {t}_{eq}\leq  {t}_{dec}=t_R$,
where, once again,  we have denoted by $t_R$ the reheating time.

\

Following the same steps as in the previous case,  now the combination of conditions 
\begin{eqnarray}
\rho_{\varphi}({t}_{dec})\leq \rho({t}_{dec}) \quad \mbox{ and } \quad \rho({t}_{eq})\sim \rho_{\varphi}({t}_{eq}),
\end{eqnarray}
leads to the constraint \cite{FL}
\begin{eqnarray}\label{bound1}
\frac{\bar{\Gamma}}{m_{\sigma}}\leq \frac{\sigma^2(t_{osc})}{3 M_{pl}^2}<1.
\end{eqnarray}

\

Since, in this case,  the reheating temperature is $T_R\sim \rho^{1/4}(t_{dec})\sim \sqrt{\sqrt{3}M_{pl}\bar{\Gamma}}$ the constraint leads to the bound
\begin{eqnarray}\label{temperature1}
T_R\leq \sqrt{\frac{m_{\sigma}}{\sqrt{3} M_{pl}}}|\sigma(t_{osc})|.
\end{eqnarray}

\

On the other hand, when the curvaton decays after its domination, the power spectrum of the curvature fluctuation is given by \cite{LW}
\begin{eqnarray}
{\mathcal P}_{\zeta}\cong \frac{1}{9\pi^2}
\frac{H_*^2}{\sigma_*^2}
\sim 2\times 10^{-9}\Longrightarrow \frac{H_*}{|\sigma_*|}\sim 4 \times 10^{-4} M_{pl},
\end{eqnarray}
and choosing, as in the previous case $H_*\sim 2\times 10^{-5} M_{pl}$,  one gets $|\sigma_*|\sim 5\times 10^{-2} M_{pl}$.  
Moreover, since the curvaton  rolls slowly before the oscillations one can take $|\sigma(t_{osc})|\sim |\sigma_*|\sim 5\times 10^{-2} M_{pl}$ which satisfies the bound
(\ref{bound1}).

\

From these values and the equation (\ref{temperature1}), one can conclude that only for curvaton  masses satisfying  
$m_{\sigma}\leq  10^{-16} M_{pl}\sim 2\times 10^2$ GeV, 
and obtaining  reheating temperatures compatible with the nucleosynthesis success.

\

Finally, dealing with the overproduction of gravitational waves, since in this case the energy density of the curvaton decreases as $a^{-3}$, the constraint (\ref{constraint})
will become
\begin{eqnarray}\label{XXX}
\Theta^{2/3}\geq 6\times 10^{-3} \left( \frac{H(0)}{M_{pl}} \right)^2,
\end{eqnarray}
where once again, $H(0)$ is the value of the Hubble parameter at the beginning of the kination epoch. The {\it heating efficiency} will approximately  be
$\Theta\cong \frac{m_{\sigma}^2\sigma_*^2}{3 H^2(0) M_{pl}^2}$, then inserting it in (\ref{XXX}) and taking $H(0)\sim 3\times 10^{-6} M_{pl}$, one obtains the constraint
$m_{\sigma}\sigma_*\geq 5\times 10^{-16} M_{pl}^2$, which  is never fulfilled. So, in that case the  gravitational waves could affect the BBN success.

\section{Conclusions}
We have studied in detail four ways to reheat  the universe in quintessence inflation via the production of particles, showing that each preheating mechanism  leads to different reheating temperatures compatibles with the Big Bang Nucleosynthesis.

\

The first one is the so-called {\it instant preheating} based in the interaction of the inflaton field with a quantum field. We have showed that for this kind of preheating the particles are produced as in the Schwinger's effect, i.e.,   the Bogoliubov coefficient 
is calculated in the same way as for the over-barrier problem in scattering theory when the
 external electric field is constant, and we have obtained  reheating temperatures of the order of $10^9$ GeV.

\

{{}
 The second mechanism consist in the production of massless particles nearly conformally coupled with gravity.
Due to the fact that the particles are nearly conformally coupled, the modes could be calculated in a perturbative way obtaining  an analytic expression of the $\beta$-Bogoliubov coefficient. For some simple models depicting phase transitions from a de Sitter phase to a kination regime we have obtained, when 
the coupling coefficient satisfies $\xi-\frac{1}{6}\sim 10^{-1}$ so  that the gravitational waves do not affect the success of the Big Bang Nucleosynthesis, 
 a reheating temperature of the order of $10^5$ GeV.}

\

{{}
The third method is the reheating via the creation of very heavy massive particles ($m_{\chi}\sim  5\times 10^{14}$ GeV) conformally coupled with gravity. 
Due to the high value of the mass, the modes could be calculated using the WKB approximation, obtaining for potentials whose first derivative is discontinuous, as the improved version of the Peebles-Vilenkin model presented in this work, analytic formulas for the energy density of the produced particles. Assuming that after the decay of these massive  particles the products thermalize instantaneously, and the overproduction of gravitational waves do not disturb the Big Bang Nucleosynthesis process,  one obtains 
a reheating temperature  around $10^2$ GeV.}

\

Finally, we have studied the reheating in quintessence inflation via the decay of the curvaton  field in very light relativistic particles, which we have assumed   thermalize instantaneously, showing that only for light masses of the curvaton field  ($m_{\sigma}\leq 10^{11}$ GeV when the curvaton is subdominant at the decay or $m_{\sigma}\leq 10^3$ GeV in the case that the energy density of the curvaton dominates the one of the inflaton at the decay) the universe is reheated at
a temperature compatible with the bounds comming from the Big Bang Nucleosynthesis.

\vspace{1cm}

{\it Acknowledgments.}   
This investigation has been supported by MINECO (Spain) grants  MTM2014-52402-C3-1-P and MTM2017-84214-C2-1-P, and  in part by the Catalan Government 2017-SGR-247.

\end{document}